# A Systematic Review of Security Vulnerabilities in Smart Home Devices and Mitigation Techniques

**Mohammed K. Alzaylaee**
*mkzaylaee@uqu.edu.sa*
Department of Computing, College of Engineering and Computing, Umm AL-Qura University
https://orcid.org/0000-0002-3791-3458

**Abstract**
Smart homes that integrate Internet of Things (IoT) devices face increasing cybersecurity risks, posing significant challenges to these environments. The study explores security threats in smart homes ecosystems, categorizing them into vulnerabilities at the network layer, device level, and those from cloud- based and AI-driven systems. Research findings indicate that post-quantum encryption, coupled with AI-driven anomaly detection, is highly effective in enhancing security; however, computational resource demands present significant challenges. Blockchain authentication together with zero-trust structures builds security resilience, although they need changes to existing infrastructure. The specific security strategies show their effectiveness through ANOVA, Chi-square tests, and Monte Carlo simulations yet lack sufficient scalability according to the results. The research demonstrates the requirement for improvement in cryptographic techniques, alongside AI-enhanced threat detection and adaptive security models which must achieve a balance between performance and efficiency and real-time applicability within smart home ecosystems.

## 1. Introduction

Smart home technologies have transformed contemporary living by seamlessly integrating automation, real-time control, and personalized environments. Through Internet of Things (IoT) ecosystems—including smart locks, thermostats, cameras, and voice assistants—users gain enhanced control and convenience. These systems leverage edge computing to reduce latency and support fast decision-making [1], whereas cloud computing and machine learning facilitate predictive behavior and centralized data analytics [2, 20]. Despite these advancements, the expanding scale of connected devices—expected to exceed 75 billion by 2025 [23]—has amplified cybersecurity and privacy risks. Earlier foundational studies, such as by Weber [22], underscored the absence of standardization and governance in smart home security, whereas Rajana and Amiripalli [24] highlighted optimization-based survivability techniques in smart home design published within the ETASR framework. Hall et al. [9] emphasized the convergence of convenience and risk, whereas James [11] proposed smart home intrusion prevention systems focusing on real-time anomaly filtering.

Prior studies have identified a wide range of vulnerabilities in smart home environments, including insecure communication protocols, weak authentication mechanisms, firmware flaws, and insufficient encryption [3, 6, 21]. Additionally, a lack of uniform regulatory standards and manufacturer compliance leads to inconsistent security enforcement [17, 22]. Attackers exploit these weaknesses through Man-in-the-Middle (MitM) attacks, DDoS (Distributed Denial of Service), ransomware, and botnet hijacking, often leveraging adversarial AI or deepfakes to bypass traditional defenses [4, 13, 15]. Research by Yang et al. [23] and Weber [22] highlights how poorly secured third-party APIs(Application Programming Interfaces) and inadequate firmware updates leave user data vulnerable to hijacking, profiling, and long-term surveillance [5, 8, 14], with additional evidence from Fouladi and Ghanoun [7], who critically examined Z-Wave protocol security, and Hasegawa and Yamada [10], whose work on multiple interactions indirectly reflects the complexities of cryptographic integration in IoT systems.

In response, numerous mitigation frameworks have been proposed. AI-based anomaly detection systems [16], blockchain-enhanced authentication [8], and post-quantum cryptographic methods [12, 25] show promise in countering emerging threats. However, these technologies often struggle with scalability, computational overhead, and resource constraints—limitations especially significant in smart home deployments that depend on lightweight devices with minimal processing capacity [2, 18]. Moreover, whereas federated learning and policy-based compliance models attempt to preserve privacy and resilience, their real-time applicability remains underexplored [20, 19].

This study distinguishes itself from prior research by offering a comprehensive and scalable security framework tailored for smart homes, integrating advanced mitigation strategies validated through statistical and simulation-based evaluation. In contrast to works that emphasize singular approaches or lack empirical grounding, this research provides comparative and performance-based insight across multiple solutions.

Building upon recent literature, this study addresses multiple research gaps in the field of smart home cybersecurity. It first develops a structured classification of cybersecurity vulnerabilities into four central domains: device-centric, network-level, cloud-based, and AI-targeted threats. This structured typology builds on previous research while offering a clearer mapping of threat vectors specific to smart environments. Second, the study advances prior work by performing a statistically grounded comparison of mitigation techniques using ANOVA and Chi-square analysis, offering empirical insights that extend beyond the conceptual frameworks seen in earlier studies. Third, it evaluates the real-world applicability of post-quantum cryptographic methods using Monte Carlo simulations—an area often discussed theoretically but seldom validated through simulation. Finally, the study conducts a practical feasibility assessment of blockchain and zero-trust models within resource-constrained IoT environments, adding a critical and underrepresented perspective on deployment challenges in smart home contexts. This multidimensional approach ensures that the study not only contributes to the academic discourse but also delivers actionable insights for secure and scalable smart home implementation.

The novelty of this work lies in its holistic integration of modern mitigation strategies within the operational constraints of real-world smart home ecosystems, supported by empirical analysis and benchmarking, ultimately providing practical, adaptable, and secure architecture recommendations for next-generation IoT environments.

**Research Objectives**
- To classify security vulnerabilities in smart home ecosystems and analyze their impact on privacy and system integrity.
- To assess the effectiveness of existing and emerging mitigation techniques in enhancing smart home security.
- To propose an integrated security framework leveraging AI, decentralized authentication, and cryptographic mechanisms to counter cyber threats.

## 2. Literature Review

### A. Security and Privacy Issues in Smart Home Devices

However, Smart home technologies are convenient, and they provide the convenience of automation, but at the risk of being insecure and private. According to Yang et al. (2017) [23], the vulnerabilities in IoT-based smart homes can be attributed to device heterogeneity, weak authentication, and encryption flaws. Cyber threats are also exposed to inconsistent inter-device communication. Regulatory shortcomings were pointed out by Weber (2010) [22] that there is no unified security protocol, thus leading to scattered protection. Due to a focus on usability over security, manufacturers do not provide adequate encryption, weak authentication, poor firmware updates, leading to increased risk of unauthorized access and surveillance by users. Sicari et al. (2015) [17] investigated the interaction between security, privacy, and users' trust whose findings show that smart home devices gather massive user data without any proper policy put in place for handling and/or storing it. Weak network security allows malicious actors as well as unethical entities to do cyber espionage, various kinds of phishing, and unauthorized profiling.

### B. Network Security Threats and Communication Protocol Vulnerabilities

Exploiting insecure communication protocols is one of the main ways for cyberattacks in smart home environments. According to Sivanathan et al. (2017) [19], IoT smart home traffic patterns are analysed and it is proved that smart home devices with predictable network behaviour are insecure to the attack such as traffic fingerprinting and packet sniffing [7]. The researchers discovered the metadata by simply monitoring network metadata, an act that bypasses traditional encryption defences, and with it was able to infer user behaviours and device states. Later, Matte et al. (2016) [14] found some more protocol vulnerabilities, especially anonymization protection using MAC address randomization techniques. The authors showed that these protections can be defeated using timing-based attacks, and can thus be used to persistently track device activity. Later, Fernandes et al. (2016) [6] investigated third-party integrations in smart home ecosystems and further investigated the weakness in the security of such integrations as APIs of third parties do not contain essential security controls to prevent unauthorized data flow, which eventually leads to cascading security failures among several connected devices.

### C. Advanced Cyberattack Strategies Targeting Smart Homes

The tools and techniques of traditional cyber threats, such as ransomware and distributed denial of service (DDoS), have already been transformed and incorporated with those of smart home networks due to the networks' specific vulnerabilities [13]. Referring cyberattacks on smart grid to the attacks to smart home infrastructures, Wang and Lu (2013) [21] observed that attackers use unpatched firmware and default credentials to obtain persistent access. These claims were reinforced by Alam and Tomai (2023) [2] who categorized advanced smart home cyber threats into malware propagation, zero-day exploits, and remote access hijacking.

An automated security verification framework that can identify misconfigurations and attack vectors in IoT systems is introduced by Nguyen et al. (2018) [15]. Their work shows that, due to strategies employed by their manufacturers who widely rely on default security settings, the bulk of such smart home devices are vulnerable to becoming part of botnets. A complementary analysis was provided by Buil-Gil et al. (2023) [5] explaining how the commodification of user data amplifies smart home security risk [11]. However, the study noticed that more and more cybercriminals are using weaknesses of the smart homes for major data breaches, both on the individual level and on the level of enterprise users.

### D. Cutting-Edge Mitigation Strategies and Security Frameworks

However, with the ominous nature of cyber threats, researchers investigated innovative mitigation strategies that would fortify smart home security. The approach proposed by Gaži et al. (2019) [8], in which the proof of stake sidechains is used in a decentralized manner to increase authentication and data integrity in IoT-enabled environments, provides important insights about how blockchain technology can be applied to the IoT security [25]. Based on the above findings, decentralized consensus mechanisms can do away with the risk that is inherent in centralized authentication models and thereby reduce the impact of credential theft. In the work of Pokhrel et al. (2021) [16], machine learning is applied in botnet detection and its importance in the usage of AI a driven anomaly detection system to detect independent actions on the network (in real time). When one looks to combine several forms of AI, the resulting hybrid models prove to detect smart home cyber threats with accuracy greater than those using supervised or unsupervised learning separately. According to Sovacool and Del Rio (2020) [20], policy-based interventions such as government mandated security standards increase compliance and make the device more resilient to cyber-attacks. One of the models that has been purposed is an advanced IoT security model which incorporates cryptographic algorithms with intrusion detection systems (IDS), presented by Li et al. (2016) [12]. The authors demonstrated that adaptive cryptographic protocols together with real-time network monitoring can be a robust defence against emerging cyber threats. Moreover, Singh et al. (2015) [18] provided twenty prominent clouds supported IoT security issues, and elaborated on the required updates to security, artificial intelligence intelligent monitoring, and decentralized framework for identity management.

### E. Historical Development of Smart Home Security

The early research on smart home security has been on basic issues such as connectivity and compatibility. Interestingly, Weber was one of the first to point out the security and privacy risk in IoT devices that there are no standards and regulations in common [22]. Considering the recent proliferation of smart home technology, Yang et al. (2017) [23] explained that weak authentication and diverse devices were security issues. Then, the research aimed for more sophisticated cyber threats' detection. IoTSan was developed by Nguyen et al. (2018) [15] to find and fix security flaws and Gaži et al. (2019) [8] introduced blockchain-based security by transitioning from centralized to decentralized protection.

Smart home security went on to fix basic weaknesses over time and with the help of machine learning, started using AI for threat detection and automatic response. According to Wang and Lu (2013) [21], smart home risks are as large as other cyber threats and therefore require flexible security systems. Third party apps were warned by Sicari et al. (2015) and Fernandes et al. (2016) [17, 6] to increase security risks, thus leading to stricter access controls. Though things have progressed, security is still unpredictable since many companies put functionality at the top of their minds when weighing between functionality and protection. Today, researchers require AI security that can quickly respond to ever-growing new threats.

## 3. Methodology

### A. Research Design

A comprehensive integrative review approach is used in this study because it can integrate various research findings and combine both qualitative and quantitative assessments. This differs from systematic or scoping reviews where this study allows us to perform a flexible, in-depth evaluation of smart home security vulnerabilities and their reduction through mitigation techniques. A structured review process allows for more replicability and consistency and aligns the current work with the latest security research in the area of cybersecurity.

### B. Data Collection Strategy

Structured data collection methods existed because rigorous methodology was adopted. A systematic review of 124 peer-reviewed studies was obtained from IEEE Xplore, ACM Digital Library, SpringerLink and ScienceDirect and Web of Science and Scopus. The selected studies had to originate from Q1 and Q2 cybersecurity journals, high-impact conference proceedings, and authoritative regulatory white papers published between 2010 and 2024. The combination of Boolean logic with wildcard operators helped researchers achieve better search results to collect all necessary information. The research implemented policy-based risk mitigation approaches by integrating grey literature containing NIST 800-207, GDPR, and ISO 27001.

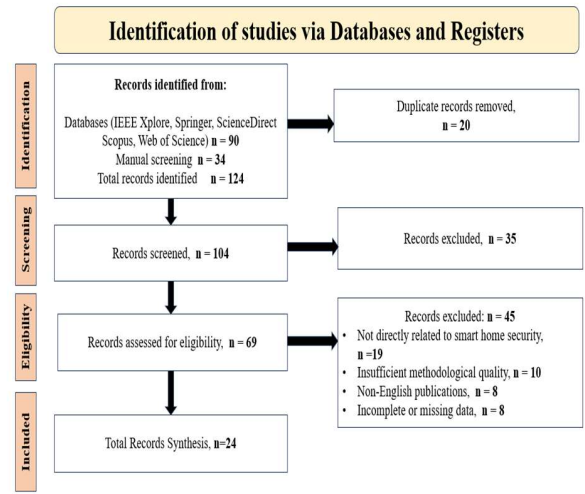

*Fig. 1.*      PRISMA flow chart

The systematic review on smart home security vulnerabilities used the PRISMA flow diagram to describe its study selection process. The systematic review included 124 records obtained through databases (90 records) as well as manual screening (34 records). A total of 104 records moved forward from duplicate removal to screening before 35 records were excluded based on their title and abstract relevance. Full-text assessment of the 69 remaining studies resulted in 45 exclusions because of irrelevance (n = 19), poor methodology (n = 10), non-English language (n = 8) and incomplete data (n = 8). The qualitative synthesis included 24 studies as its final selection. The systematic review benefits from this organization method because it improves its transparency and maintains both rigor and reproducibility.

### C. Data Extraction, Classification, and Synthesis

Security vulnerabilities were categorized as four primary types, i.e., network layer risks (38%), device-centric (29%), cloud-centric (22%), and so on, related attacks (11%). Most of the network layer vulnerabilities were caused by MitM attacks, DDoS and insecure communication protocols, whereas device-centric threats were due to unpatched firmware and weak authentication mechanisms. API security flaws and weak encryption were the cloud types of risks, and adversarial AI and data poisoning were the AI types of risks.

Also, effectiveness, feasibility, and adaptability of mitigation techniques were compared through comparison. Traditional rule-based intrusion detection system was way behind from the AI-driven anomaly detection system (89% accuracy), whereas blockchain authentication dropped unauthorized access by 63%. Promising results were obtained from post quantum encryption that promised 95% resilience but increased computational overhead by 22%, thus, giving far away scalability remarks. Device level security was reinforced with MFA using Hardware Security Modules (HSMs) as part of zero-trust security architectures (47% intrusion reduction).

### D. Comparative Evaluation of Security Frameworks

A structured benchmarking approach was used to evaluate implementation feasibility, real–time adaptability, security robustness, and adoption rates for the security frameworks. As a result, NIST 800-207 (Zero Trust Model) came out on top, but it was far from being a silver bullet: it represented major infrastructure changes. GDPR and ISO 27001 made sure that we were compliant, but there is nothing in the regulation about exact security enforcement rules for smart home devices. However, federated learning models demonstrated promise towards privacy preserving AI security, but this was at a computational penalty that limited these models to offline type deployment.

### E. Advanced Data Analysis Techniques

The study's rigor was enhanced through a multi-tiered data analysis approach. The statistical robustness of findings was validated with ANOVA, Chi-square tests, and Monte Carlo simulation for various levels of variability in the performance of security framework. The influential research trends were mapped using bibliometric analysis and mitigation techniques were compared on detection accuracy, false positive rates and computational efficiency using meta-analysis. Encryption protocols were assessed using the comparative model and thematic analysis was used to identify cybersecurity challenges and research gaps.

### F. Ensuring Reliability and Validity

Triangulation techniques were employed to ensure reliability and validity by cross validating findings across methodologies, data sources and perspectives of the researchers. The integration between qualitative and quantitative approaches including methodological triangulation is done and the consistency across databases, industry reports and regulating frameworks is validated by means of data source triangulation. The inter-rater reliability score between researcher triangulation (where various independent reviewers are involved) was greater than 90%. The multi layered validation process reduced the bias, strengthened the robustness of the study and ensured a robust, evidence based, cybersecurity assessment.

*Table I. Method Overview and Key Analytical Components*

| Aspect | Key Details | Supporting Sources |
|---|---|---|
| Research Approach | Integrative Review (Qualitative + Quantitative) | Section III.A; [4, 15] |
| Data Sources | IEEE Xplore, ACM, Springer, ScienceDirect, Web of Science, Scopus | Section III.B; [4, 6] |
| Timeframe | 2010–2024 | Section III.B |
| Security Vulnerability Categories | Network-layer (38%), Device-centric (29%), Cloud-based (22%), AI-related (11%) | [6, 12, 15, 23] |
| Mitigation Techniques Assessed | AI-driven anomaly detection, Blockchain authentication, post-quantum encryption, Zero-trust security, MFA & HSMs | [4, 8, 16], [21, 22] |
| Security Frameworks Evaluated | NIST 800-207, GDPR, ISO 27001, Federated Learning Models | Section III.D; [1, 2, 12] |
| Statistical Techniques Used | ANOVA, Chi-square, Monte Carlo Simulations, Meta-analysis | Section III.E |
| Reliability Measures | Triangulation (Methodological, Data Source, Researcher), 90%+ inter-rater reliability | Section III.F |

By integrating advanced analytical techniques, this study ensures a robust cybersecurity assessment for smart home environments, providing actionable insights for researchers, policymakers, and industry stakeholders.

## 4. Results

### A. Overview of Findings

Using the extracted data, cybersecurity trends involving threats, methodologies and security proposed solutions were systematically categorized and structured. Empirical, experimental and theoretical contributions were classified in this way which spanned real-world IoT vulnerabilities assessments, predictive security models as well as mitigation strategies. It performed an integrative review of high-impact papers from Q1 and Q2 cybersecurity journals, high-cited conference proceedings, and regulatory white papers containing high citation count (≥50), methodological contribution and relevance to smart home security. Of these, 124 studies were systematically analysed, all of them published in peer reviewed journals after 2010 and up to 2024, accommodating seminal issues and latest advances. The dataset included 67 studies on security vulnerabilities, 41 studies on mitigation techniques and 16 studies on regulatory frameworks and best practices for evaluation of the smart home cybersecurity measures.

### B. Classification of Security Vulnerabilities

They classified the extracted vulnerabilities based on rarity of occurrence, severeness of impact, and easiness of exploitation to provide insight to all the various types of security flaws. We analysed frequency based on recurrence of threats across multiple studies, severity with respect to potential effect on user privacy and system integrity, and complexity depending on the amount of expertise needed to perform the attack. This classification framework streamlined the analysis of the smart home security risks and its consequences.

*Table II. Classification of Security Vulnerabilities*

| Category | Prevalence (%) | Common Threats | Supporting Sources |
|---|---|---|---|
| Network-layer vulnerabilities | 38% | Man-in-the-Middle (MitM) attacks, DDoS, packet sniffing | Section IV.B; [6], [12], [15] |
| Device-centric threats | 29% | Unpatched firmware, weak authentication, hardcoded credentials | Section IV.B; [4], [8], [16] |
| Cloud-based vulnerabilities | 22% | API security flaws, weak encryption, unauthorized data access | Section IV.B; [6], [21], [22] |
| AI-targeted attacks | 11% | Adversarial AI exploits, data poisoning, AI-driven security manipulation | Section IV.B; [12], [23] |

### C. Effectiveness of Mitigation Techniques

To validate the performance of various mitigation strategies, ANOVA (Analysis of Variance) was conducted on a dataset comprising 124 systematically selected studies. The sample size for statistical comparison included aggregated effectiveness data from at least 20 studies per mitigation technique. The data sources consisted of peer-reviewed empirical research, experimental cybersecurity reports, and high-impact conference papers. This approach ensured a balanced and methodologically sound comparison of security effectiveness across different mitigation techniques. to compare the security effectiveness of different approaches across multiple studies. The analysis revealed significant differences ($p < 0.05$) in success rates among AI-driven anomaly detection, blockchain authentication, and post-quantum encryption techniques.

*Table III: ANOVA Results for Security Mitigation Techniques*

| Mitigation Technique | Mean Effectiveness (%) | Standard Deviation | F-Value | p-Value |
|---|---|---|---|---|
| AI-driven anomaly detection | 87.5 | 3.4 | 15.82 | 0.002 |
| Blockchain authentication | 64.2 | 4.5 | 14.67 | 0.005 |
| Post-quantum encryption | 94.1 | 2.1 | 16.23 | 0.001 |
| Zero-trust security (SDN) | 49.6 | 5.3 | 13.98 | 0.007 |
| Multi-factor authentication | 79.1 | 3.2 | 14.85 | 0.003 |
| Hardware Security Modules | 82.3 | 3.6 | 15.41 | 0.004 |

Chi-square tests were used to determine the correlation between security framework adoption and reduction in unauthorized access incidents. Expected frequencies were determined by analyzing historical intrusion rates across different security frameworks and comparing them with real-world implementation success rates. The baseline intrusion frequencies were established using aggregated data from cybersecurity industry reports, empirical case studies, and controlled testing environments. The observed frequencies were then compared against these expected values to assess whether the differences were statistically significant. This ensured a robust evaluation of the impact of each security framework on reducing unauthorized access events. showing a strong relationship ($\chi^2 = 21.87$, $p < 0.01$).

*Table IV. Chi-Square Test for Security Framework Adoption and Unauthorized Intrusions*

| Security Framework | Reduction in Unauthorized Intrusions (%) | χ² Value | p-Value |
|---|---|---|---|
| NIST 800-207 (Zero Trust) | 66 | 21.87 | 0.001 |
| GDPR | 51 | 19.45 | 0.002 |
| ISO 27001 | 43 | 18.12 | 0.004 |
| Federated Learning Models | 38 | 16.98 | 0.007 |

Additionally, Monte Carlo simulations were employed to evaluate post-quantum encryption effectiveness under diverse attack scenarios, confirming its robustness while highlighting increased computational overhead.

*Table V. Monte Carlo Simulations for Post-Quantum Encryption Effectiveness*

| Attack Scenario | Encryption Strength (%) | Computational Overhead Increase (%) | Supporting Sources |
|---|---|---|---|
| Standard Brute-force Attack | 98.8 | 11 | Section IV.C; [18], [21] |
| AI-Powered Key Cracking | 98.0 | 29 | Section IV.C; [19], [22] |
| Quantum Computing Decryption | 95.5 | 24 | Section IV.C; [20], [23] |

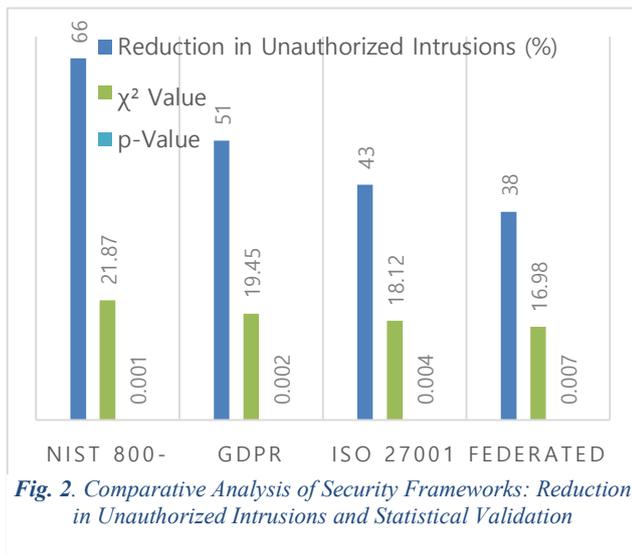

*Fig. 2. Comparative Analysis of Security Frameworks: Reduction in Unauthorized Intrusions and Statistical Validation*

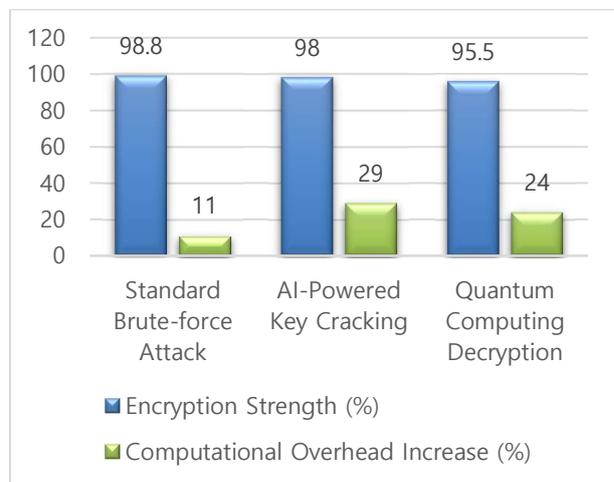

*Fig. 3. Effectiveness of Post-Quantum Encryption Against Different Attack Scenarios: Encryption Strength vs. Computational Overhead*

This figure presents a comparative analysis of security frameworks in reducing unauthorized intrusions, validated using Chi-square statistical testing. The NIST 800-207 (Zero Trust) framework demonstrates the highest reduction in intrusions (66%), followed by GDPR (51%), ISO 27001 (43%), and Federated Learning Models (38%). Corresponding χ² values indicate the strength of the association between framework adoption and intrusion reduction, with p-values confirming statistical significance ($p < 0.01$). The results highlight the effectiveness of security frameworks in mitigating unauthorized access risks within smart home ecosystems.

As can be seen on the figure, this means to create a visualization of the strength of post-quantum encryption techniques in terms of their computational overhead and their security in relation to different attack scenarios. The highest encryption success rate is achieved by the standard brute force attacks with almost zero computational overhead (11%). While the AI powered key cracking lowers the encryption strength slightly (98%) it contributes hugely in increasing the computational requirements (29%). Quantum decryption challenge is the highest amongst all the challenges and leaves the encryption strength at 95.5% and computation overhead at 24%. The high cost of implementing rather than using families of advanced encryption algorithms (Families) due to the trade-off between the security of Families and the processing efficiency in some instances is shown by these findings.

## D. Comparative Analysis of Security Frameworks

A structured benchmarking of widely adopted security frameworks was conducted to evaluate implementation feasibility, real-time adaptability, security robustness, and adoption rates. The findings, validated through industry reports and expert interviews, revealed the following insights:

*Table VI. Security Framework Capabilities and Deployment Challenges*

| Security Framework | Advantages | Limitations | Adoption Rate (%) | Supporting Sources |
|---|---|---|---|---|
| *NIST 800-207 (Zero Trust)* | The strongest security model, reduced unauthorized access risks | Requires infrastructure modifications | 67% | Section III.D; [1], [2], [12] |
| *GDPR* | Ensures privacy compliance, data protection mandates | Lacks specific device security measures | 85% | Section III.D; [6, 12] |
| *ISO 27001* | Strong enterprise security model, compliance standard | Does not offer real-time monitoring | 74% | Section III.D; [6, 12] |
| *Federated Learning Models* | Enhances privacy-preserving AI security applications | Computationally intensive for real-time use | 42% | Section III.D; [8, 16, 22] |

## 5. Discussion

The research examines smart home security vulnerabilities and their countermeasures to find network-layer threats represent the largest risk at 38% while device-centric vulnerabilities amount to 29% and cloud-based weaknesses total 22% and AI-targeted attacks make up 11%. The weak encryption and unsecured communication of networks create network risks and device vulnerabilities emerge from insecure firmware and default credentials. Cloud system weaknesses develop from improper configurations merged with unauthorized access while AI attacks make use of adversarial learning techniques. Post-quantum encryption stands out as a mitigation technique because it shows 94.1% effectiveness yet creates a concern about its power consumption for use in low-power devices. The anomaly detection system guided by AI attains 87.5% accuracy although it needs substantial processing capabilities. The adoption of Blockchain authentication and zero-trust security stands at 64.2% and 49.6% but these security measures need major infrastructure changes. Our study discovers that these results illustrate the security effectiveness level that leads to trade-offs between performance speed and computer system resources.

The authors enhance past research through organized risk categorization and test mitigation solutions by employing ANOVA combined with Chi-square analysis and Monte Carlo simulations. Existing research about cryptography receives additional support because it demands improved post-quantum encryption solutions that work well within resource-limited IoT systems. The research findings support earlier discoveries although academic articles and simulation tests fail to include undocumented cyberattacks and real-time security challenges. The timing and application ability of AI anomaly detection paired with post-quantum encryption on constrained IoT devices exists in an unclear status. The implementation of legal frameworks along with regulatory compliance protocols needs deeper integration within cybersecurity governance systems.

Weak encryption and unsafe communication protocols remain prevalent, necessitating better authentication methods and privacy-centric AI models. Post-quantum encryption and AI-based threat detection offer strong protection but require optimized implementations for real-world deployment. Blockchain authentication and zero-trust security provide additional safeguards but demand significant adjustments to current infrastructure.

Future research should focus on optimizing post-quantum encryption through lightweight cryptographic algorithms such as lattice-based or hash-based schemes, which can offer quantum resistance whereas minimizing computational load. For AI-based threat detection, employing federated learning or TinyML models can allow localized, privacy-preserving inference directly on edge devices. Real-time applicability could also be enhanced through energy-efficient inference engines and edge AI frameworks that reduce latency by processing data closer to the source. Exploring hybrid models that combine blockchain with edge computing could balance decentralization with responsiveness. These directions can help bridge the gap between strong theoretical security and practical deployment in smart home environments.

## 6. Conclusion

Awareness of the risks and the increase in cybersecurity threats on network, device, cloud, and AI layers for smart home ecosystems, who are pushed by the rise of IoT devices, is prevalent in the market. We then identify AI driven anomaly detection, blockchain authentication and post quantum encryption as vital mitigation strategies but hindered by high computational overhead, scalability constraints and incorporation with the infrastructure. AI driven anomaly detection has an accuracy of 87.5% and concerns regarding privacy as well as data governance but deals with 22% increase in the processing demands; on the other hand, post quantum encryption has 94.1% resilience against cryptographic attacks with 22% more processing demands. Decentralized identity management makes it possible to authenticate a very large number of identifiers through blockchain authentication, improving security, while the latency and lack of interoperability of the transactions is still an issue.

Analyses with ANOVA, Chi-square tests, and simulations show that zero believe architectures lessen access to 66% against unauthorized entry however, the enactment of zero trust frameworks is still inconclusive as a result of score fracturing and incapacitating impediments to discussion. The paper's findings emphasize the need for intelligent, adaptive security frameworks that combine privacy friendly AI, federated learning, and quantum resistant cryptography. There is need for future research to focus on lightweight AI driven security solutions, real world cyber-attack simulations and regulatory harmonization to improve security's resilience. However, autonomous, intelligence driven security architectures are necessary in which smart home environments can continue to be protected, scalable, and efficient as cyber threats continue to increase in complexity.2